\documentclass[onecolumn,prd,aps,12pt]{revtex4}%
\usepackage{amsmath}
\usepackage{amsfonts}
\usepackage{amssymb}
\usepackage{graphicx}%
\providecommand{\U}[1]{\protect\rule{.1in}{.1in}}
\begin{document}
\title{Modified Maxwell equations from CPT-even Lorentz violation with Minimum Length}
\author{T. Prud\^encio$^{1}$\footnote{email: prudencio.thiago@ufma.br, thprudencio@gmail.com}, L. S. Amorim$^{2}%
$\footnote{email: lesamorim@gmail.com }, H. Belich$^{2}$\footnote{email: belichjr@gmail.com}, H. L. C. Louzada$^{2}$\footnote{email: haofisica@bol.com.br.}}
\affiliation{$^{1}${\small Coordination of Science \& Technology (CCCT-BICT), Federal
University of Maranh\~ao (UFMA), 65080-805, S\~ao Lu\'is-MA, Brazil.}}
\affiliation{$^{2}${\small Departamento de F\'{\i}sica e Qu\'{\i}mica, Federal University 
of Esp\'{\i}rito Santo (UFES), 29060-900, Vit\'{o}ria, ES, Brazil.}}
\affiliation{}
\date{\today}

\begin{abstract}
Here we discuss the presence of CPT-even Lorentz violation (LV) in the
presence of a deformed Heisenberg algebra that leads to a minimum length (ML).
We consider the case of a Maxwell lagrangian modified by the presence of a
$K_{F}$ CPT-even LV theory and ML. We then derive a set of modified Maxwell
equations in the cases of LV and ML and only ML. We verified that in the case
of electromagnetic waves in the vacuum the presence of ML does not change the
consequences of LV. On the other hand, in a material media ML changes the
whole set of equations that can lead to important effects with respect to the
usual equations. We also considered the more general case including LV and the
modified equations in terms of matter fields. We then derived the refractive
index as a function of the matter fields depending on LV and ML, and in
particular we showed the behaviour of the refractive index with respect to the
non-commutative parameter.

\end{abstract}
\maketitle



\section{Introduction}

Despite the great success of Standard Model (SM) in describing through field
theory the regimes that unifies weak and electromagnetic interactions, SM has
clear limitations concercening regimes of unification in the Planck era,
mainly related to the presence of gravity. The interest in investigating the
physics beyond SM has been increased with the need of understanding the
problem of Dark Matter. The regimes of interaction between the dark and
non-dark sectors, can induce the detection of a weak fifth force. This was
tested investigating anomalous decays of an excited state of $^{8}Be$
\cite{irvin}. Also, we have the unbalance between matter-antimatter that has
not been clarified by the SM
\cite{Axion,Axion1,Axion2,Axion3,Axion4,Axion5,Axion6}.

Investigating the physics beyond SM, Kosteleck\'{y} and Samuel \cite{sam}
proposed a new regime where spontaneous violation of symmetry occurs through
non-scalar fields based on string field theory, leading to a vacuum with a
tensor nature . A consistent description of fluctuations around this new
vacuum is obtained if the components of the background field are constant, and
by the fact that the minimum in the the background not represented by a
scalar, and consequently Lorentz symmetry is spontaneously broken \cite{ens}.

This extension SM by Lorentz symmetry violation (LV) has been considered for
fields that belong to a more fundamental theory, which may induce the
spontaneous violation of Lorentz symmetry based on a specific potential. It is
worth mentioning that this extension of SM keeps the gauge invariance, the
conservation of energy and momentum and the covariance under observer
rotations and boosts, where this extension is called as the Standard Model
Extension (SME) \cite{col,coll2}. In this context, it is well-known that the
presence of terms that violate the Lorentz symmetry imposes at least one
privileged direction in the spacetime. In recent decades, studies of the
Lorentz violation (LV) have been made in several branches of physics
\cite{h1,h2,e1,e2,e3,e4,e5,e6,e6a,e7,rasb1,w,tensor1,tensor2,geom1,geom2,geom3,bb2,bb4,book,l1,g1,g2,g3}%
. The LV has been investigated in two major scenarios proposed: spontaneous
Lorentz symmetry violation (SLV) caused by a tensorial background treated
above, and the breaking made by generalization of uncertainty principle - the
non-commutative geometry.

On the other hand the proposal of noncommutative geometry was developed in
1980 by A. Connes \cite{33} and it was realized that the non-commutative
geometry would be a scheme to extend the standard model in several ways
\cite{necra}. In the 90s the proposal appears naturally in the context of
string theory \cite{34, 44}. In this way we may obtain an effective theory
describing scenarios in string theory whose in the low energy limit is reduced
to a known physical theory.

Noncommutative geometry also appear in a condensed matter context as an
effective theory that describes the electron in a two-dimensional surface
attached to a strong magnetic field. This effective theory describes the
Quantum Hall Effect. The electron would be trapped in the lowest Landau levels
and presents the Hall conductance in $e^{2}/{\hbar}$ units \cite{56}. The
effect of non-commutativity could be tested for instance in an hydrogen atom,
one of the simplest quantum systems that allows theoretical predictions and
experimental verifications of high accuracy \cite{SG}. There are many papers
where the energy spectrum of the hydrogen atom in the presence minimum length
is calculated \cite{FB,SB,RA}, some of which have divergences in levels
$s$($n=1$) \cite{SB}.

A possible way to explore the implementation of noncommutatives theories is by
the deformation of the Heisenberg algebra. In particular, a modified
Heisenberg algebra is achieved by adding certain small corrections to the
canonical commutation relations, as shown by A. Kempf and contributors
\cite{K1,K2,K3,K4,K5}, to the minimum uncertainty in the position measurement,
$\delta x_{0},$ called minimum length. The existence of this minimum length
was also suggested by quantum gravity and string theory \cite{DJ,MM,EW}.

Recently, Quesne and Tkachuk have introduced a Lorentz covariant deformed
algebra that describes a quantized $D+1$-dimentional \cite{QT,QT2}, it is
given by the following generalized commutation relations:
%
\begin{align}
\left[  X^{\mu},P^{\nu}\right]    & =-i\hbar\beta\left[  (\frac{1}{\beta
}-P_{\rho}P^{\rho})g^{\mu\nu}-\frac{\beta^{^{\prime}}}{\beta}P^{\mu}P^{\nu
}\right]  \label{es1}\\
\left[  P^{\mu},P^{\nu}\right]    & =0\\
\left[  X^{\mu},X^{\nu}\right]    & =i\hbar\beta\frac{\lbrack(2-\frac
{\beta^{^{\prime}}}{\beta})-(2+\frac{\beta^{^{\prime}}}{\beta})P_{\rho}%
P^{\rho}](P^{\mu}X^{\nu}-P^{\nu}X^{\mu})}{(\frac{1}{\beta}-P_{\rho}P^{\rho})},
\end{align}
where $\mu,\nu,\rho=0,1,\cdots,D$, $\quad{g_{\mu\nu}}=g^{\mu\nu}%
=diag(1,-1,-1,\cdots,-1)$, $\quad\beta$ and $\beta^{^{\prime}}$ are
deformation parameters, and we suppose $\beta,\beta^{^{\prime}}>0$. From
uncertaint relation we conclude that the minimum length (ML) is
%
\[
\left(  \delta X^{i}\right)  _{0}=\hbar\beta\sqrt{(D+\frac{\beta^{^{\prime}}%
}{\beta})\left[  \frac{1}{\beta}-\left\langle (P^{0})^{2}\right\rangle
\right]  }\quad,\quad\forall i\in\left\{  1,\cdots,D\right\}  .
\]
A algebra representation \cite{VM} that satisfies (\ref{es1}) in the first
order in $\beta,\beta^{^{\prime}}$\ is given by:
%
\begin{align}
X^{\mu}  & =x^{\mu}-\beta\frac{(1-\frac{\beta^{^{\prime}}}{2\beta})}{2}%
(x^{\mu}p_{\rho}p^{\rho}+p_{\rho}p^{\rho}x^{\mu}),\label{es3}\\
P^{\mu}  & =\frac{\left(  \frac{2}{\beta^{^{\prime}}}-p_{\rho}p^{\rho}\right)
}{2}\beta^{^{\prime}}p^{\mu},
\end{align}
where $x^{\mu}$ and $p^{\mu}=i\hbar{\partial}^{\mu}$ are the position and
momentum operators. Particular cases are achieved for $\beta^{^{\prime}%
}=2\beta$ and $\beta^{^{\prime}}=\beta$ the Quesne-Tkachuk algebra is
simplified.
%

Here we investigate the scenario of anisotropy in polarized electromagnetic
waves generated by the presence of a Lorentz symmetry breaking tensor $\left(
K_{F}\right)  _{\mu\nu\alpha\beta}$ \cite{klink1}, which appears in Standard
Model Extended (SME) in the CPT-even gauge sector \cite{klink1},
\cite{klink2}, and in the presence of a minimum length provided by a
non-commutative structure of a Heisenberg algebra. The effects of these
anisotropies in the nature of vacuum polarized electromagnetic waves is then discussed.

The structure of this paper is the following: {Sections II we presents our
modified electromagnetism with minimum length (ML) in the CPT-even gauge
sector LV. In section III, discuss the effect of the ML in the set of Maxwell
equations. In the section IV, we derive the set of Maxwell equations and
Matter fields in the presence of ML in the CPT-even gauge LV. Section V, we
derive the matter fields in the Fourier transformed space and obtain the
corresponding refractive index in the presence of LV and ML. Finally, in
section VI, we address our conclusions. }

\section{CPT-even gauge sector with minimum length}

A CPT-even gauge sector of SME can be described by the following lagrangean
\cite{casana}
\begin{equation}
{\mathcal{L}}_{2N}=-\frac{1}{4\mu_{0}}\left(  F_{\mu\nu}F^{\mu\nu}
-(K_{F})_{\mu\nu\kappa\lambda}F^{\mu\nu}F^{\kappa\lambda} \right) -A_{\mu
}J^{\mu} ,
\end{equation}
where $(K_{F})_{\mu\nu\kappa\lambda}$ is a tensor with non-dimensional and
renormalizable coupling tensor responsible by LV, whose symmetries are the
same as the Riemann tensor and the double trace vanishes, i. e.,
\begin{align}
&  (K_{F})_{\mu\nu\kappa\lambda} =-(K_{F})_{\nu\mu\kappa\lambda},\quad
(K_{F})_{\mu\nu\kappa\lambda}=-(K_{F})_{\mu\nu\lambda\kappa},\quad(K_{F}%
)_{\mu\nu\kappa\lambda}=(K_{F})_{\kappa\lambda\mu\nu};\\
&  (K_{F})_{\mu\nu\kappa\lambda}+(K_{F})_{\mu\kappa\lambda\nu}+(K_{F}%
)_{\mu\lambda\nu\kappa}=0,\label{ib2}\\
&  {(K_{F})^{\mu\nu}}_{\mu\nu}=0. \label{dtn}%
\end{align}
We then write this lagrangean in the presence of the minimum length
(\ref{es3}), i.e.,
\begin{align}
x^{\mu}\rightarrow X^{\mu}  &  = x^{\mu},\\
\partial^{\mu}\rightarrow\nabla^{\mu}  &  =(1+\beta\hbar^{2}\Box)\partial
^{\mu} ,\nonumber
\end{align}
where $\Box=\partial_{\mu}\partial^{\mu}$. Neglecting terms $O(2)$ in $\beta$,
we obtain
\begin{align}
{\mathcal{L}}_{2NM}  & = -\frac{1}{4\mu_{0}}\left( F_{\mu\nu}F^{\mu\nu}
-(K_{F})_{\mu\nu\kappa\lambda}F^{\mu\nu}F^{\kappa\lambda}\right)  -\frac
{\beta\hbar^{2}}{2\mu_{0}}\left( F_{\mu\nu}\Box F^{\mu\nu} - (K_{F})_{\mu
\nu\kappa\lambda}F^{\mu\nu}\Box F^{\kappa\lambda}\right)  - A_{\mu}J^{\nu
}\nonumber
\end{align}
or equivalent
\begin{align}
{\mathcal{L}}_{2NM}  & = {\mathcal{L}}_{2N} -\frac{\beta\hbar^{2}}{2\mu_{0}%
}\left( F_{\mu\nu}\Box F^{\mu\nu} - (K_{F})_{\mu\nu\kappa\lambda}F^{\mu\nu
}\Box F^{\kappa\lambda}\right) .
\end{align}
This will lead to the following equations
\begin{align}
\label{em1}\left(  1+2\beta\hbar^{2}\Box\right)  \left[  \partial_{\nu}%
F^{\nu\mu}-{(K_{F})^{\mu\nu}}_{\rho\phi}\partial_{\nu}F^{\rho\phi}\right]   &
=\mu_{0}J^{\mu}.
\end{align}
A parametrization of this theory is described in \cite{Kostelec3,Kostelec2},
where 19 independent components of $(K_{F})$ are described in terms of four
matrices $3\times3$, named: $(\kappa_{DE}), (\kappa_{HB}), (\kappa_{DB})$ e
$(\kappa_{HE})$. These component relations are given by
\begin{equation}
(\kappa_{DE})^{jk}=-2(K_{F})^{0j0k},\quad(\kappa_{HB})^{jk}=\frac{1}%
{2}\epsilon^{jpq}\epsilon^{klm}(K_{F})^{pqlm},
\end{equation}
\begin{equation}
(\kappa_{DB})^{jk}=-(\kappa_{HE})^{kj}=\epsilon^{kpq}(K_{F})^{0jpq},
\end{equation}
from which we see that $(\kappa_{DE})$ and $(\kappa_{HB})$ are symmetric,
while $(\kappa_{DB})$ is not. We then take $\mu=0$, in (\ref{em1}), leading
the following modified Gauss law
\begin{equation}
\left( 1+2\beta\hbar^{2}\Box\right)  \left[  \partial_{i}E^{i}+(\kappa
_{DE})_{lj}\partial_{l}E^{j}+c(\kappa_{DB})_{lk}\partial_{l}B_{k}\right]
=\frac{\rho}{\epsilon_{0}}. \label{Lç}%
\end{equation}
Taking $\mu=i$, we obtain the modified Amp\`ere-Maxwell law
\begin{align}
&  {}\left(  1+2\beta\hbar^{2}\Box\right)  [-\partial_{t}E^{i}/{c^{2}%
}+\epsilon_{ijk}\partial_{j}B^{k}-(\kappa_{DE})_{ij}\partial_{t}E^{j}/{c^{2}%
}+(\kappa_{DB})_{ik}\partial_{t}B^{k}/{c}+\nonumber\\
&  {}{(\kappa_{HB})_{jk}\epsilon_{jip}\partial_{p}B^{k}+ \epsilon_{ipk}%
(\kappa_{DB})_{mk}\partial_{p}E^{m}}/{c}]=\mu_{0}J^{i}.
\end{align}
In a vector form, these equations are written as
\begin{equation}
\left( 1+2\beta\hbar^{2}\Box\right)  \left[  \nabla\cdot\mathbf{E} +\left(
\kappa_{DE}\cdot\nabla\right) \cdot\mathbf{E}+c\left( \kappa_{DB}\cdot
\nabla\right) \cdot\mathbf{B}\right]  =\frac{\rho}{\epsilon_{0}}. \label{lk1}%
\end{equation}
\begin{align}
&  {}\left(  1+2\beta\hbar^{2}\Box\right)  [-\partial_{t}\mathbf{E}/{c^{2}}+
\nabla\times\mathbf{B}-\kappa_{DE}\cdot\partial_{t}\mathbf{E}/{c^{2}}%
+\kappa_{DB}\cdot\partial_{t}\mathbf{B}/{c}+\nonumber\\
&  + \nabla\times(\kappa_{HB}\cdot\mathbf{B}) + \nabla\times(\kappa_{DB}%
\cdot\mathbf{E})/{c}]=\mu_{0}\mathbf{J}. \label{lk2}%
\end{align}
In the vacuum ($J^{\mu}=0$) the minimum length modifies Gauss and
Amp\`ere-Maxwell laws by the presence of a global factor resulting from
$(1+2\beta\hbar^{2}\Box)$. The dispersion relation will furnish the modes
${p^{0}}^{2}= |\mathbf{p}|^{2}+ 1/2\beta$ and the particle mass relation
$m=1/2\sqrt{\beta}c$. We conclude that for the CPT-even gauge sector of SME,
the presence of SLV and minimum length are independent effects, i. e., even
for an electrodynamics without LV there is a massive pole resulting from
non-commutativity \cite{SKM}.

\section{Maxwell equations changed by minimum length}

Let us consider the whole set of Maxwell equations
\begin{align}
\nabla\cdot\mathbf{E} & = \frac{\rho}{\varepsilon_{0}},\\
\nabla\cdot\mathbf{B} & = 0,\\
\nabla\times\mathbf{E} & = -\frac{\partial\mathbf{B}}{\partial t},\\
\nabla\times\mathbf{B} & = \mu_{0}\mathbf{J} + \mu_{0}\varepsilon_{0}%
\frac{\partial\mathbf{E}}{\partial t},
\end{align}
The presence of a minimum length transfomation due to non-commutativity
\begin{align}
\nabla & \rightarrow (1+\beta\hbar^{2}\Box)\nabla\\
\frac{\partial}{\partial t}  & \rightarrow(1+\beta\hbar^{2}\Box)\frac
{\partial}{\partial t}%
\end{align}
will lead to the following modified Maxwell equations
\begin{align}
(1+\beta\hbar^{2}\Box)\nabla\cdot\mathbf{E} & = \frac{\rho}{\varepsilon_{0}%
},\label{s31}\\
(1+\beta\hbar^{2}\Box)\nabla\cdot\mathbf{B} & = 0,\\
(1+\beta\hbar^{2}\Box)\nabla\times\mathbf{E} & = -(1+\beta\hbar^{2}\Box
)\frac{\partial\mathbf{B}}{\partial t},\\
(1+\beta\hbar^{2}\Box)\nabla\times\mathbf{B} & = \mu_{0}\mathbf{J} + \mu
_{0}\varepsilon_{0}(1+\beta\hbar^{2}\Box)\frac{\partial\mathbf{E}}{\partial
t},\label{s34}%
\end{align}
Consequently, electromagnetic waves in the presence of a minimum length will
remain the same as if this type of commutativity was not present. The only
distinction is in the presence of a source terms, eqs. (\ref{s31}) and
(\ref{s34}). In this case, as usual, let us split the current density in terms
of free, polarization and magnetization contributions
\begin{align}
\mathbf{J} = \mathbf{J}_{f} + \mathbf{J}_{\mathbf{P}} + \mathbf{J}%
_{\mathbf{M}},
\end{align}
and the charge density in terms of free and polarization terms
\begin{align}
\rho=\rho_{f}+\rho_{\mathbf{P}},
\end{align}
where the material media has associated electric polarization $\mathbf{P}$ and
magnetization $\mathbf{M}$, with corresponding definitions
\begin{align}
\rho_{\mathbf{P}} & = -\nabla\cdot\mathbf{P}\\
\mathbf{J}_{\mathbf{P}} & = \frac{\partial\mathbf{P} }{\partial t}\\
\mathbf{J}_{\mathbf{M}} & = \nabla\times\mathbf{M}.
\end{align}
where Applying a divergent $\nabla\cdot$ in eq. (\ref{s34}),
\begin{align}
\nabla\cdot(1+\beta\hbar^{2}\Box)\nabla\times\mathbf{B} & = \mu_{0}\nabla
\cdot\mathbf{J} + \mu_{0}\varepsilon_{0}\nabla\cdot(1+\beta\hbar^{2}\Box
)\frac{\partial\mathbf{E}}{\partial t},
\end{align}
and taking into account eq. (\ref{s31}), we have the validity of continuity
equation
\begin{align}
\nabla\cdot\mathbf{J} + \frac{\partial\rho}{\partial t} = 0,
\end{align}
and also for the electric polarization
\begin{align}
\nabla\cdot\mathbf{J}_{\mathbf{P}} + \frac{\partial\rho_{\mathbf{P}}}{\partial
t} = 0.
\end{align}
Thus, the charge density can be separated in the free part $\rho_{f}$ and the
part depending on polarization, and the current density has a contribution due
to free contributions $J_{f}$, polarization and magnetization, as given by
\begin{align}
\mathbf{J} = \mathbf{J}_{f} + \mathbf{J}_{\mathbf{P}} + \mathbf{J}%
_{\mathbf{M}}.
\end{align}
The first modified Maxwell equation leads to the following generalization
involving a matter field
\begin{align}
\nabla\cdot\mathbf{D}_{\beta}=\rho_{f}\label{dbeta}%
\end{align}
where we have a generalized response $\mathbf{D}$ to the material media
\begin{align}
\mathbf{D}_{\beta} & =\left( \varepsilon_{0}(1+\beta\hbar^{2}\Box)\mathbf{E} +
\mathbf{P}\right) \nonumber\\
& = \mathbf{D} + \beta\hbar^{2}\Box\mathbf{E}%
\end{align}
On the other hand, we then have
\begin{align}
(1+\beta\hbar^{2}\Box)\nabla\times\mathbf{B} & = \mu_{0}\left(  \mathbf{J}_{f}
+ \frac{\partial\mathbf{P}}{\partial t} + \nabla\times\mathbf{M}\right)  +
\mu_{0}\varepsilon_{0}(1+\beta\hbar^{2}\Box)\frac{\partial\mathbf{E}}{\partial
t},
\end{align}
where we also have a generalized response $\mathbf{H}$ to the material media
\begin{align}
\mathbf{H}_{\beta} & = \frac{(1+\beta\hbar^{2}\Box)}{\mu_{0}}\mathbf{B}
-\mathbf{M}\nonumber\\
& = \mathbf{H} + \frac{\beta\hbar^{2}\Box}{\mu_{0}}\mathbf{B}.
\end{align}
and the generalized equation
\begin{align}
\nabla\times\mathbf{H}_{\beta}  & =\mathbf{J}_{f} + \frac{\partial
\mathbf{D}_{\beta}}{\partial t}.\label{hbeta}%
\end{align}
Using the constitutive relation for polarization $\mathbf{P}= \varepsilon
_{0}\chi_{e}\mathbf{E}$ and the generalized one for magnetization
\begin{align}
\mathbf{M} & = \chi_{m,\beta}\mathbf{H}_{\beta},
\end{align}
we then have generalized relations for the fields in material media depending
on operator coming from ML
\begin{align}
\mathbf{D}_{\beta} & = \varepsilon_{0}\left( 1+\beta\hbar^{2}\Box+ \chi
_{e}\right) \mathbf{E},\\
\mathbf{H}_{\beta} & = \frac{1}{(1+\chi_{m,\beta})\mu_{0}}(1+\beta\hbar
^{2}\Box)\mathbf{B}.
\end{align}

We then have generalized equations involving matter field in the material
media
\begin{align}
(1+\beta\hbar^{2}\Box)\nabla\cdot\mathbf{D}_{\beta} & = \left( 1+\beta
\hbar^{2}\Box+ \chi_{e}\right) \rho,\label{s3wf}\\
\nabla\cdot\mathbf{H}_{\beta} & = 0,\\
(1+\beta\hbar^{2}\Box)\nabla\times\mathbf{D}_{\beta}  & = -\left( 1+\beta
\hbar^{2}\Box+ \chi_{e}\right) (1+\chi_{m,\beta})\varepsilon_{0}\mu_{0}%
\frac{\partial\mathbf{H}_{\beta}}{\partial t},\\
(1+\chi_{m,\beta})\mu_{0}\nabla\times\mathbf{H}_{\beta} & = \mu_{0}\mathbf{J}
+ \mu_{0}(1+\beta\hbar^{2}\Box)\frac{\partial}{\partial t}\left( 1+\beta
\hbar^{2}\Box+ \chi_{e}\right) ^{-1}\mathbf{D}_{\beta},\label{s4wf}%
\end{align}
where the equations (\ref{dbeta}) and (\ref{hbeta}) are simplest forms of eqs.
(\ref{s3wf}) and (\ref{s4wf}), when expressed in terms of free charge and
current densities.

\section{Maxwell equations changed by minimum length and Lorentz violation}

Let us now consider the equations (\ref{lk1}) and (\ref{lk2}), using
$\beta^{\prime}=2\beta$,
\begin{equation}
\left( 1+\beta^{\prime2}\Box\right)  \left[  \nabla\cdot\mathbf{E} +\left(
\kappa_{DE}\cdot\nabla\right) \cdot\mathbf{E}+c\left( \kappa_{DB}\cdot
\nabla\right) \cdot\mathbf{B}\right]  =\frac{\rho}{\epsilon_{0}}. \label{lkk1}%
\end{equation}
\begin{align}
&  {}\left(  1+\beta^{\prime2}\Box\right)  [-\partial_{t}\mathbf{E}/{c^{2}}+
\nabla\times\mathbf{B}-\kappa_{DE}\cdot\partial_{t}\mathbf{E}/{c^{2}}%
+\kappa_{DB}\cdot\partial_{t}\mathbf{B}/{c}+\nonumber\\
&  + \nabla\times(\kappa_{HB}\cdot\mathbf{B}) + \nabla\times(\kappa_{DB}%
\cdot\mathbf{E})/{c}]=\mu_{0}\mathbf{J}, \label{lkk2}%
\end{align}
that, with the equations
\begin{align}
(1+\beta\hbar^{2}\Box)\nabla\cdot\mathbf{B} & = 0,\\
(1+\beta\hbar^{2}\Box)\nabla\times\mathbf{E} & = -(1+\beta\hbar^{2}\Box
)\frac{\partial\mathbf{B}}{\partial t},
\end{align}
form the set of modified Maxwell equations with CPT-even LV and minimum length.

Taking into account the previous definitions, we have
\begin{align}
\left( 1+\beta^{\prime2}\Box\right)  \left[  \nabla+\left( \kappa_{DE}%
\cdot\nabla\right)  \right] \cdot\mathbf{D}_{\beta^{\prime}}  & = \left(
1+\beta^{\prime2}\Box+ \chi_{e}\right) \left[ \rho-\frac{(1+\chi
_{m,\beta^{\prime}})}{c}\left( \kappa_{DB}\cdot\nabla\right) \cdot
\mathbf{H}_{\beta^{\prime}}\right] ,\nonumber\\
& \\
\mu_{0}(1+\chi_{m,\beta})\nabla\cdot\mathbf{H}_{\beta} & = 0,\\
(1+\beta\hbar^{2}\Box)\nabla\times\mathbf{D}_{\beta} & = -\frac{1}{c^{2}%
}\left( 1+\beta\hbar^{2}\Box+ \chi_{e}\right) (1+\chi_{m,\beta})\frac
{\partial\mathbf{H}_{\beta}}{\partial t},
\end{align}
\begin{align}
\label{lkk8i} &  \frac{1}{c^{2}}\left( 1+\beta\hbar^{2}\Box+ \chi_{e}\right)
(1+\chi_{m,\beta^{\prime}})\left[ \nabla\times(\mathbf{H}_{\beta^{\prime}}
+\kappa_{HB}\cdot\mathbf{H}_{\beta^{\prime}}) + \frac{1}{c}\kappa_{DB}%
\cdot\partial_{t}\mathbf{H}_{\beta^{\prime}}\right] \nonumber\\
& =\frac{1}{c^{2}}\left( 1+\beta\hbar^{2}\Box+ \chi_{e}\right) \mathbf{J} +
\frac{1}{c^{2}}\left(  1+\beta^{\prime2}\Box\right) \left[ \partial_{t}\left(
\mathbf{D}_{\beta} + \kappa_{DE}\cdot\mathbf{D}_{\beta}\right) -c\nabla
\times(\kappa_{DB}\cdot\mathbf{D}_{\beta})\right] \nonumber\\
\end{align}
We can consider the case where
\begin{align}
\kappa_{DB}\cdot\mathbf{D}_{\beta} & =\kappa_{DE}\cdot\mathbf{D}_{\beta}=0,\\
\kappa_{HB}\cdot\mathbf{H}_{\beta^{\prime}} & = 0,\\
\kappa_{DB}\cdot\partial_{t}\mathbf{H}_{\beta^{\prime}} & = 0.
\end{align}
that will reduce eq. (\ref{lkk8i}) to the following
\begin{align}
\left( 1+\beta\hbar^{2}\Box+ \chi_{e}\right) (1+\chi_{m,\beta^{\prime}}%
)\nabla\times\mathbf{H}_{\beta^{\prime}}  & =\left( 1+\beta\hbar^{2}\Box+
\chi_{e}\right) \mathbf{J} + \left(  1+\beta^{\prime2}\Box\right) \partial_{t}
\mathbf{D}_{\beta}\nonumber\\
\end{align}

\section{Matter fields in the Fourier transformed space}

\begin{figure}[h]
\centering
\includegraphics[scale=0.4]{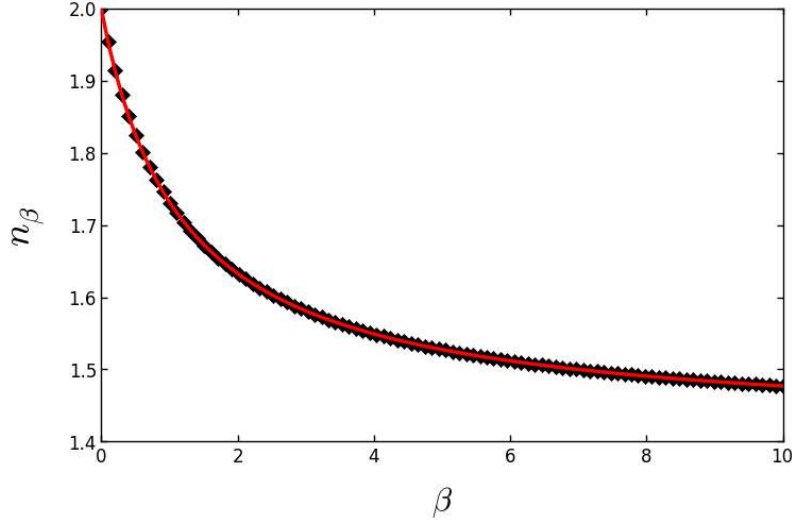}
\caption{(Color online) Refractive index as a function of the non-commutative
parameter $\beta$.}%
\label{tghiiz12}%
\end{figure}

Taking the generalized form of a matter field in the scenario of LV with ML,
in the Fourier transformed space, we have
\begin{align}
\mathbf{D}_{\beta}(\mathbf{p},\omega=p_{0}) & = \varepsilon_{0}\left(
1+\beta\hbar^{2}p_{\mu}p^{\mu} + \chi_{e}\right) \mathbf{E}(\mathbf{p}%
,\omega=p_{0}),\\
\mathbf{H}_{\beta}(\mathbf{p},\omega=p_{0})  & = \frac{1}{(1+\chi_{m,\beta
})\mu_{0}}(1+\beta\hbar^{2}p_{\mu}p^{\mu})\mathbf{B}(\mathbf{p},\omega=p_{0}).
\end{align}
We then have the corresponding
\begin{align}
\varepsilon_{\beta}(\mathbf{p},\omega=p_{0}) & =\varepsilon_{0}\left(
1+\beta\hbar^{2}p_{\mu}p^{\mu} + \chi_{e}\right) \\
\mu_{\beta}(\mathbf{p},\omega=p_{0}) & =\frac{(1+\chi_{m,\beta})\mu_{0}%
}{(1+\beta\hbar^{2}p_{\mu}p^{\mu})}%
\end{align}
The refractive index associated to the material is then given by
\begin{align}
n_{\beta} & = \sqrt{\left( 1+\beta\hbar^{2}p_{\mu}p^{\mu} + \chi_{e}\right)
\frac{(1+\chi_{m,\beta})}{(1+\beta\hbar^{2}p_{\mu}p^{\mu})}}%
\end{align}
In particular, we display in the figure \ref{tghiiz12}, for $\hbar^{2}=1,
\chi_{m,\beta}=1, \chi_{e}=1, p_{\mu}p^{\mu}=1$, the behavior of the
refractive index as a function of $\beta$ in the ML for this CPT-even LV
scenario with ML. We also have that, in the absence of ML, the refractive
index is modified by the presence of a LV encapsulated in the generalized
$\chi_{e}$ and $\chi_{m,\beta}$ in the presence of LV. In the limits
$\chi_{m,\beta}<<1$, $\chi_{e}<<1$, the refractive index is also reduced to a
$\pm1$, where the $-1$ corresponds to a metamaterial behaviour.

\section{Concluding remarks}

We have considered a CPT-even gauge sector of SME in the presence of a
deformed Heisenberg algebra with minimum length (ML) . In this scenario, we
derived a set of modified Maxwell equations. We conclude that the usual
effects of modified electromagnetic waves does not change in the presence of a
ML. However, in a material media, the effects of the deformed algebra in the
set of Maxwell equation could be verified even in absence of LV. We derived
both sets of Modified Maxwell equations in material media, i.e., with and
without LV. In particular, we derived the dielectric functions in the Fourier
transformed space and derived the refractive index in the presence of LV and
ML; finally we showed how the refractive index is related to the
non-commutative parameter.

These results are important aspects for the tests with SME and
non-commutativity, in particular, in condensed matter scenarios that could
verify the tensors of SME and the parameters of ML, with experimental tests.

\textbf{Ackowledgements}

The authors acknowledge the supports by CNPq, CAPES, FAPES and
FAPEMA-UNIVERSAL-01401/16 (Brazil).

\end{document}